\documentstyle[epsfig,12pt,worldsci]{article}
\begin{document}
\title{PROPERTIES OF THE ${\bf B_{c}}$ SYSTEM}
\author{Lewis P. Fulcher \\
{\em Department of Physics and Astronomy, Bowling Green State University \\
    Bowling Green, Ohio 43403, USA\\ email: fulcher@newton.bgsu.edu \\}}
\maketitle
\setlength{\baselineskip}{2.6ex}

\vspace{0.7cm}
\begin{abstract}

I summarize a comprehensive account of the energies, splittings and
electromagnetic decays of the low-lying states of the bottom charmed meson
system. Richardson's potential is used to include running coupling constant
effects in the central potential, and the full radiative one-loop expressions
derived by Pantaleone, Tye and Ng are used for the spin-dependent potentials.
Our predicted result for the ground state energy is $6286^{+15}_{-6} \: MeV$.
The ground state lifetime is found to be $\tau_{B_{c}} = 0.36 \pm 0.05 \: ps $.
Both of these numbers are consistent with the measured values recently
announced by the CDF collaboration at Fermilab.
\end{abstract}
\vspace{0.7cm}

\section{Introduction}

The CDF collaboration \cite{ab98} at the Fermilab Tevatron has recently reported
the discovery of $B_{c}$ mesons. Their announced value for the mass of the
ground state is $6400 \pm 390 \pm 130 \: MeV $,	and their value for the 
ground state lifetime is $\tau_{B_{c}} = 0.46^{+0.18}_{-0.16} \pm 0.03 \: ps $.
This state should be one of a number of relatively stable states below the 
threshold for the emission of B and D mesons, and thus I want to present
some results from a comprehensive calculation \cite{fu98} of 
the energies, splittings and electromagnetic decay rates
of the low-lying states of the $B_{c}$ system. 

In 1991 Kwong and Rosner \cite{kw91} predicted the masses of the two 
lowest states of this system from an empirical mass formula and a logartihmic
potential. They obtained a range for the ground state, $ 6194 \leq M_{B_{c}}
\leq 6292 \: MeV $. Eichten and Quigg \cite{ei94} gave a more comprehensive
account of the properties of the $B_{c}$ system. Using several potential models
to obtain error estimates, they quoted $ 6258 \pm 20 \: MeV $ for the ground
state energy. In anticipation of the experimental interest in
the $B_{c}$ system, there have been a number
of other potential model calculations \cite{ge95}.

\section{Richardson's potential and the spin-dependent potentials}

Richardson's potential \cite{ri79} contains a confining potential and a
short range piece, 
\begin{equation}
\label{rip}
V(r) = A r - \: \frac{8\pi \, f(\Lambda r ) }{(33 - 2 n_{f})r} \;, \;
f(t) = \frac{4}{\pi } \int_{0}^{\infty } \frac{sin\: tx}{x} \left[
\frac{1}{\ln (1 + x^{2})} - \frac{1}{x^{2}} \right] dx,
\end{equation}
where the string constant will be treated as a free parameter instead of
a function of $\Lambda$ in order to achieve a separation of the long and
short distance parameters \cite{fu91}.

To specify the spin-dependent potentials, we will use Pantaleone, Tye
and Ng's (PTN) generalization \cite{pa86} of the Eichten-Feinberg formalism.
Thus 
\small
\begin{eqnarray*}
\label{sd}
V_{SD} & = & \frac{{\bf L} \cdot {\bf S}_{1} } { 2 m_{1}^{2} } 
       \left[\frac{1}{r} \frac{d}{d\,r} (E(r) - A r + 2 V_{1}) 
	+ 2 V_{5} \right]
       + \frac{{\bf L} \cdot {\bf S}_{2} } { 2 m_{2}^{2} }
	\left[ \frac{1}{r} \frac{d}{d\,r} (E(r) - Ar + 2 V_{1})
	- 2 V_{5} \right] \nonumber \\
  & & + \frac{{\bf L} \cdot ({\bf S}_{1} + {\bf S}_{2}) } { m_{1} m_{2} } 
	\frac{1}{r} \frac{d\, V_{2}}{d\, r}
+ \frac{{\bf L} \cdot ({\bf S}_{1} - {\bf S}_{2}) }{m_{1} m_{2}} V_{5}
+ \frac{\bf (S_{\rm 1} \cdot \hat{r} \,S_{\rm 2} \cdot \hat{r}
    - \frac{\rm 1}{\rm 3} S_{\rm 1}\cdot S_{\rm 2})}{m_{1} m_{2}} V_{3} 
+ \frac{ {\bf S}_{1} \cdot {\bf S}_{2} } {3 m_{1} m_{2} } V_{4}, 
\end{eqnarray*}
\normalsize
where the component potentials E, $V_{1}$, $V_{2}$, $V_{3}$, $V_{4}$ and
$V_{5}$, include both the tree-level and the full radiative one-loop
contributions. These may be
expressed in terms of the strong coupling constant
$\alpha_{\bar{S}}$, the renormalization scale $\mu $ and the constituent
masses. As an example, we list the second spin-orbit potential,             

\small
\begin{equation}
V_{2}(r) = - \frac{ 4 \alpha_{\bar{S}}}{3 r} \left[ 1 + \frac{\alpha_{\bar{S}}}
      { 6\pi }\left[ (33 - 2 n_{f} ) (\ln \mu r + \gamma_{e} ) + \frac{39}{2}
      - \frac{5 n_{f} }{3} - 9 ( \ln \sqrt{m_{1} m_{2}} r + \gamma_{E} )
      \right] \right],
\end{equation}
\normalsize
where $\gamma_{E} $ is Euler's constant.
All of the PTN component potentials are given in the modified minimal
subtraction scheme. The central potential parameters are assumed to be
flavor independent. These parameters and the constituent masses are determined 
from the low-lying states of charmonium and the upsilon system. Their values are
\begin{equation}
A=0.152 \: {\rm GeV^{2}}, \:  \Lambda = 0.431 \: {\rm GeV}, \:
m_{b} =4.889 \: {\rm GeV }, \: m_{c} =1.476 \: {\rm GeV},
\end{equation}
and $n_{f}$ is taken to be 3.
For the upsilon system the value of the coupling constant $\alpha_{\bar S }
= 0.30, $ and the value of the renormalization scale $\mu = 1.95 \: {\rm GeV}$.
For charmonium these values are $\alpha_{\bar S } = 0.486$ and $\mu = 0.80 \:
{\rm GeV} $. In both cases the universal QCD scale $\Lambda_{QCD}=0.190\: GeV$,
a value consistent with other determinations \cite{pa96}.
These parameters give an average deviation of 4.3 MeV for the upsilon states
below the continuum threshold and an average deviation of 19.9 MeV for the
corresponding charmonium states.
The one-loop potentials give a substantial advantage in simultaneously
accounting for the fine structure and the hyperfine structure of charmonium
and the upsilon system. Thus, it is reasonable to expect they should 
produce better results for the spin-dependent effects in the $B_{c}$
system.

\section{Results and conclusions}

Since the central potential parameters are assumed to be strictly flavor
independent and the constituent masses are not allowed to run, the only
decision that one has to make to treat the $B_{c}$ system is to choose
a value for $\alpha_{\bar{S}} $. We choose $\alpha_{\bar{S}} = 0.393 $, the
average of the values for charmonium and the upsilon system. Then 
preserving the value of the universal QCD scale requires that $\mu = 1.12 \:
GeV $. Our calculated results for the three lowest levels of the $B_{c}$
system are shown in Figure~1, where they are compared 
with earlier potential model calculations \cite{ei94,ge95} and recent
lattice calculations. \cite{da96}
The 190 MeV error associated with uncertainty of the overall energy scale of
the lattice calculation is not shown. In Figure~1 FUII98 denotes the
one-loop results and the FUI98 results are based on tree-level expressions
for the spin-dependent potentials, and one can see substantial differences
for both the S and the P states. The FUI98 results are very close to the
results of Eichten and Quigg, as expected.

\begin{figure}
\begin{center}
\psfig{figure=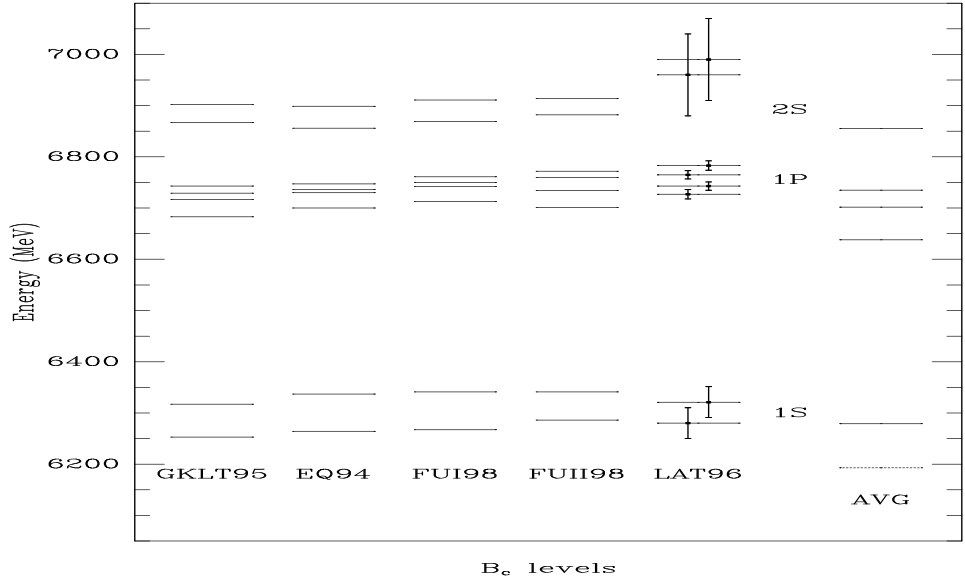,height=3.0in,width=5.0in}
\caption{\footnotesize Low-lying levels of the $B_{c}$ system. The averages are
        computed from charmonium and upsilon levels.}
\end{center}
\end{figure}

An important difference between the spectrum of the $B_{c}$ system and those
of charmonium and the upsilon system arises because the total spin 
${\bf S } = {\bf S}_{1} + {\bf S}_{2} $ no longer commutes with the 
Hamiltonian. Thus one must compute a mass-mixing matrix for P states, and it 
convenient to use the j-j coupling scheme \cite{fu98,ei94} for this purpose. 
After computing the matrix representatives of each of the spin-dependent
operators in the spin-dependent potential, one finds that 
the $J=2$ states and the $J=0$ state  are the same as in the L-S basis, but
that the $J=1$ states are mixtures. Our result for the lowest $J=1$ state is 
\begin{equation}
\psi_{1m}(1^{+}) = 0.118 \psi_{1m}(3/2,1/2) + 0.993 \psi_{1m}(1/2,1/2) ,
\end{equation}
which is very near the j-j coupling limit, a result consistent
with the lattice calculations \cite{da96}. In this regard we differ from
Eichten and Quigg \cite{ei94}, whose tree-level results were much closer
to the L-S limit. This difference is reflected in the character of the
spectrum of photons following the dipole radiative decays of the 1P state. 
Assuming an equal distribution of initial populations, our
results for the photon spectrum of this state contains 6 distinct lines,
as shown in Figure~2, instead of the 4 one would expect in the L-S
limit.

\begin{figure}
\begin{center}
\psfig{figure=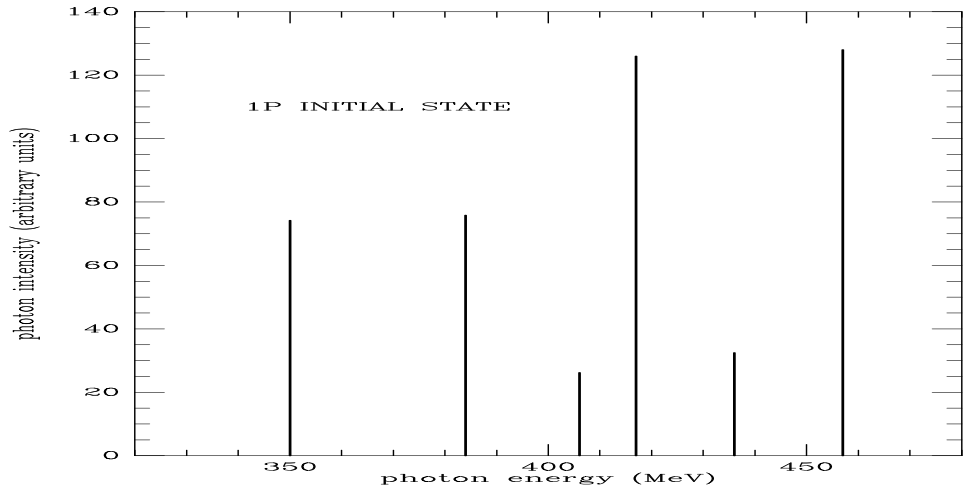,height=2.5in,width=5.0in}
\caption{\footnotesize Photon spectrum of the 1P state.}
\end{center}
\end{figure}

To estimate the uncertainty of our approach, we have allowed $\alpha_{S} $
to vary from its charmonium value (0.486) to its upsilon value (0.30). Holding
$\Lambda_{QCD} = 0.190 \: GeV $ allows an error estimate of the ground state,
that is,  $M_{B_{c}} = 6286^{+15}_{-6} \: MeV.$
In order to determine the lifetime of the ground state, we have considered
$\bar{b} $-quark decay, where the charmed quark is a spectator, and c-quark
decay, where the the antibottom quark is a spectator, and the annihilation
channels reached by the $B_{c}$ decay constant. Thus
$\Gamma_{tot}(B_{c}) =  1.85 \pm 0.24 \times 10^{-3} \: eV,$
where most of the error arises from the uncertainty \cite{pa96} in
the Kobayashi-Maskawa matrix element $V_{bc}$.


\vskip 1 cm
\thebibliography{References}

\bibitem{ab98}
CDF collaboration, F. Abe {\em et al.}, submitted to Phys. Rev. D, available
as lanl eprint hep-ex/9804014.

\bibitem{fu98} submitted to Phys. Rev. D, available as a lanl eprint
hep-ph/9806444.

\bibitem{kw91}
W. Kwong and J. Rosner, Phys. Rev. D{\bf 44}, 212 (1991).

\bibitem{ei94}
E. Eichten and C. Quigg, Phys. Rev. D{\bf 49}, 5845 (1994).

\bibitem{ge95}
S. Gershtein, V. Kiselev, A. Likhoded and A. Tkabladze, Phys. Rev. D {\bf 51},
3613 (1995);
Y. Chen and Y. Kuang, Phys. Rev. D {\bf 46}, 1165 (1992);
V. Galkin, A. Mishurov and R. Faustov, Sov. J. Nucl. Phys. {\bf 55}, 1207
(1992).

\bibitem{ri79} J. Richardson, Phys. Lett. {\bf 82B}, 272 (1979).

\bibitem{fu91} L. Fulcher, Phys. Rev. D {\bf 44}, 2079 (1991).

\bibitem{pa86} J. Pantaleone, S. Tye and Y. Ng, Phys. Rev. D {\bf 33}, 777
(1986).

\bibitem{pa96} Particle Data Group, R. Barnett {\em et al.}, Phys. Rev. D
{\bf 54}, 1 (1996).

\bibitem{da96} 
C. Davies, K. Hornbostel, G. Lepage, A. Lidsey, J. Shigemitsu and J. Sloan,
Phys. Lett. B {\bf 382}, 131 (1996).
\end{document}